%% file: main.tex
\documentclass[sigconf,screen]{acmart}

\AtBeginDocument{%
  }

\usepackage{graphicx}
\usepackage{microtype}
\usepackage{balance}

\usepackage{booktabs}
\usepackage{multirow}
\usepackage{listings}

\lstdefinestyle{promptblock}{
  basicstyle=\ttfamily\tiny,
  breaklines=true,
  breakatwhitespace=false,
  columns=fullflexible,
  keepspaces=true,
  frame=single,
  framerule=0.2pt,
  rulecolor=\color{black!25},
  backgroundcolor=\color{black!2},
  xleftmargin=0.5em,
  xrightmargin=0.5em,
  aboveskip=0.6em,
  belowskip=0.6em,
  captionpos=t,
  literate=
    {—}{{---}}1
    {–}{{--}}1
    {→}{{$\to$}}1
    {×}{{$\times$}}1
    {“}{{``}}1
    {”}{{''}}1
    {’}{{'}}1
}

\setcopyright{cc}
\setcctype{by-nc-nd}
\acmDOI{10.1145/3844177.3844556}
\acmYear{2026}
\copyrightyear{2026}
\acmISBN{979-8-4007-3001-6/2026/10}
\acmConference[Harness4GenUI '26]{Proceedings of the 1st International
  Workshop on Harness Engineering for Generative UI: Everything Evolves,
  Something Endures}{October 12--16, 2026}{Munich, Germany}
\acmBooktitle{Proceedings of the 1st International Workshop on Harness
  Engineering for Generative UI: Everything Evolves, Something Endures
  (Harness4GenUI '26), October 12--16, 2026, Munich, Germany}
\acmSubmissionID{asews26harnessgenuimain-p5-p}
\received{2026-07-26}
\received[accepted]{2026-08-22}

\begin{document}

\title{Toward Frontier-Quality Declarative UI Generation at Small-Model Cost}

\author{Yingxiang Yang}
\authornote{Both authors contributed equally to this research.}
\correspondingauthor
\orcid{0009-0005-0158-1841}
\affiliation{%
  \institution{Amazon}
  \city{Seattle}
  \country{USA}
}
\email{yayingxi@amazon.com}

\author{Weihang Xiao}
\authornotemark[1]
\authornote{Work done while at Amazon.}
\orcid{0009-0009-6203-4138}
\affiliation{%
  \institution{Amazon}
  \city{Seattle}
  \country{USA}
}
\email{wx228@cornell.edu}

\author{Ben Bullough}
\authornotemark[2]
\orcid{0009-0002-7716-1606}
\affiliation{%
  \institution{Amazon}
  \city{Seattle}
  \country{USA}
}
\email{ben.bullough@gmail.com}

\author{Tushar Deshpande}
\orcid{0009-0007-7183-0756}
\affiliation{%
  \institution{Amazon}
  \city{Seattle}
  \country{USA}
}
\email{tsdesh@amazon.com}

\author{Niresh Agarwal}
\orcid{0009-0005-7116-551X}
\affiliation{%
  \institution{Amazon}
  \city{Seattle}
  \country{USA}
}
\email{nirea@amazon.com}

\renewcommand{\shortauthors}{Yang, Xiao, Bullough, Deshpande, and Agarwal}

\input{sections/abstract}

%% CCS concepts and keywords are required for articles over two pages.
%% Replace the placeholder CCS codes via https://dl.acm.org/ccs/ccs.cfm



\maketitle

\input{sections/introduction}
\input{sections/related_work}
\input{sections/methodology}
\input{sections/evaluation}
\input{sections/discussion}
\input{sections/conclusion}

\balance

% ---- Bibliography ----
\bibliographystyle{ACM-Reference-Format}
\bibliography{references}

% ---- Appendix (sections are lettered in the ACM appendix) ----
% Dropped to fit the 8-page limit; re-enable both lines to restore the
% extra qualitative renders in sections/appendix.tex.
% \appendix
% \input{sections/appendix}

\end{document}

%% file: sections/abstract.tex
\begin{abstract}
Declarative UI protocols such as A2UI let applications generate interactive UIs by selecting pre-built components from a catalog and binding their props to application data, rather than emitting frontend code from scratch. This contract is attractive for production systems because of safety and consistency. An open question is: can low-latency and low-cost small models achieve the required quality for A2UI-based UI generation? To answer this, we systematically study three controllable design choices for catalog-conditioned A2UI generation: supervised fine-tuning (SFT) data construction method, model size, and component-catalog size. Across two React/TypeScript domains and four base checkpoints spanning two model families (Qwen 3.5 0.8B/2B/4B; SmolLM 3B), we find: (i) a 4B fine-tuned student recovers $\sim$98\% of teacher semantic quality and $\sim$97\% of teacher visual quality at more than an order of magnitude lower cost than frontier API calls; (ii) both augmented strategies (Perturbed-catalog and Constrained-GT) Pareto-dominate the unaugmented Full-catalog baseline, while specializing on different axes; (iii) even small models can handle and benefit from relatively large component catalog size. We distill these results into practitioner-facing trade-offs and deployment recommendations across the three design choices.
\end{abstract}

%% file: sections/introduction.tex
\section{Introduction}

Conversational AI is shifting from text answers to interactive interfaces. UI generation capability has become an important dimension in LLM code generation evaluation, with live arenas \cite{lmarena2025webdev,arena2026code} and dedicated benchmarks \cite{xiao2025designbench,jung2025uibench} now measuring how well frontier models build interactive web applications and component-based UIs. The strategic question is no longer whether to render a UI on demand, but how to deliver these surfaces at deployable quality, latency, and cost. Declarative UI protocols such as A2UI \cite{a2ui2025} are an emerging contract for this setting: the model emits a structured description selecting components from a catalog and binding their props to data paths; a client renderer turns this into a live interface. Currently frontier models clear the quality bar but raise real concerns about latency and cost; off-the-shelf small models miss quality outright (in our setup, base Qwen 3.5 4B \cite{qwen2026qwen35} parses valid A2UI on 62\% of inputs and the 2B base on 3\%). Supervised fine-tuning is one natural remedy, but the practical trade-offs across SFT data curation, model size, and component catalog have not been characterized for declarative UI generation in the literature.

\paragraph{Our study.} We close that gap with end-to-end ablations across all three knobs, evaluated under in-distribution (ID) and stricter out-of-distribution (OOD) catalog-shift conditions. We use a task-management application as the primary domain (86-component React/TypeScript catalog, $\sim$1{,}500 utterance-and-data pairs) and replicate the main findings on a separate cloud-management console UI domain. We organize the study around four research questions. \textbf{(RQ1)} Which SFT data-construction strategy works best, and for which operational requirement? We compare three, used under these names throughout and defined in \S\ref{sec:sft_strategies}: \emph{Full-catalog} (canonical baseline), \emph{Perturbed-catalog} (catalog shuffling, dropout, renaming), and \emph{Constrained-GT} (simpler targets under a component budget, milder renaming). \textbf{(RQ2)} What model size is needed? We measure 0.8B, 2B, and 4B Qwen students and place them against frontier APIs on cost and latency. \textbf{(RQ3)} How does catalog size shape student quality? We sweep from $\sim$10 to $\sim$100 components at every student size. \textbf{(RQ4)} Do the findings replicate on a different base model and domain? We repeat the comparison with SmolLM 3B on a cloud-management console.

\paragraph{Main findings.} (1) A 4B SFT student reaches 3.45/5 semantic and 3.09/5 visual on task-management from a 1.76/1.57 base, recovering $\sim$98\% of teacher semantic and $\sim$97\% of teacher visual quality, at more than an order of magnitude lower per-query cost than calling SOTA commercial models through an API (\S\ref{sec:rq2}). (2) Both augmented strategies Pareto-dominate the unaugmented Full-catalog baseline, and they trade off against each other on a quality--reliability axis: Perturbed-catalog wins average semantic/visual quality, while Constrained-GT makes the trained model more reliable (higher data binding and path resolution accuracy). (3) Catalog richness scales student quality monotonically from $\sim$10 to $\sim$100 components: even small student models benefit instead of being overwhelmed by component catalog richness. (4) The strategy ranking replicates on a different base model and UI domain (SmolLM 3B on cloud-management), where Constrained-GT's data-binding advantage again holds and its semantic quality matches Perturbed-catalog.

%% file: sections/related_work.tex
\section{Related Work}

\paragraph{LLM-driven UI generation.}
LLMs have been applied to UI generation across three regimes that differ in what the model emits. Raw frontend code from a visual or natural-language specification is the most-studied regime: pix2code \cite{beltramelli2018pix2code} mapped a single screenshot to GUI markup, and multimodal LLMs now produce HTML/CSS at production quality \cite{si2024design2code,laurencon2024websight,yun2024web2code}. Framework-specific UI code (React, Vue, Angular) and the broader front-end workflow (generation, edit, repair) are the focus of recent benchmarks such as DesignBench \cite{xiao2025designbench}. Declarative, JSON-based UI protocols, the regime our paper studies, restrict the model's output to a structured description that selects pre-built components from an approved catalog and binds props to data paths, which the client renderer turns into a native UI without executing arbitrary code \cite{a2ui2025}. To our knowledge, no public benchmark has characterized small-model SFT for the declarative regime; our study fills that gap.

\paragraph{Evaluating generative UI\@.}
Evaluation of LLM-generated UI typically combines three signals: (1) parse and render success on the target framework or markup; (2) visual similarity between the rendered output and a reference screenshot \cite{si2024design2code,laurencon2024websight,yun2024web2code,xiao2025designbench}; and (3) judge-based quality scoring on the rendered surface using LLM judges \cite{zheng2023judging,liu2023geval,kim2024prometheus}. For declarative protocols, structured-validity contracts (component-catalog membership, executable data bindings, parse correctness) gate downstream usefulness, and judge-based semantic and visual scoring on the rendered surface complements the structured checks. We adopt this combined regime in our evaluation (\S\ref{sec:methodology}).

\paragraph{LLM-as-judge.}
Structured-generation quality cannot be captured by surface similarity alone, so we combine parse/render checks with LLM judges \cite{zheng2023judging,liu2023geval,kim2024prometheus,wang2023selfinstruct,xu2023wizardlm,gudibande2023false} alongside executable metrics (path resolution, catalog hallucination), mitigating known judge biases \cite{panickssery2024llmevaluators} via a separate human-annotation calibration (\S\ref{sec:judge_calibration}). Our OOD evaluation tests expanded-catalog generalization: the model must select and bind components introduced after SFT, not merely handle paraphrased user requests.

%% file: sections/methodology.tex
\section{Methodology}
\label{sec:methodology}

\begin{figure*}[t]
\centering
\includegraphics[width=\textwidth]{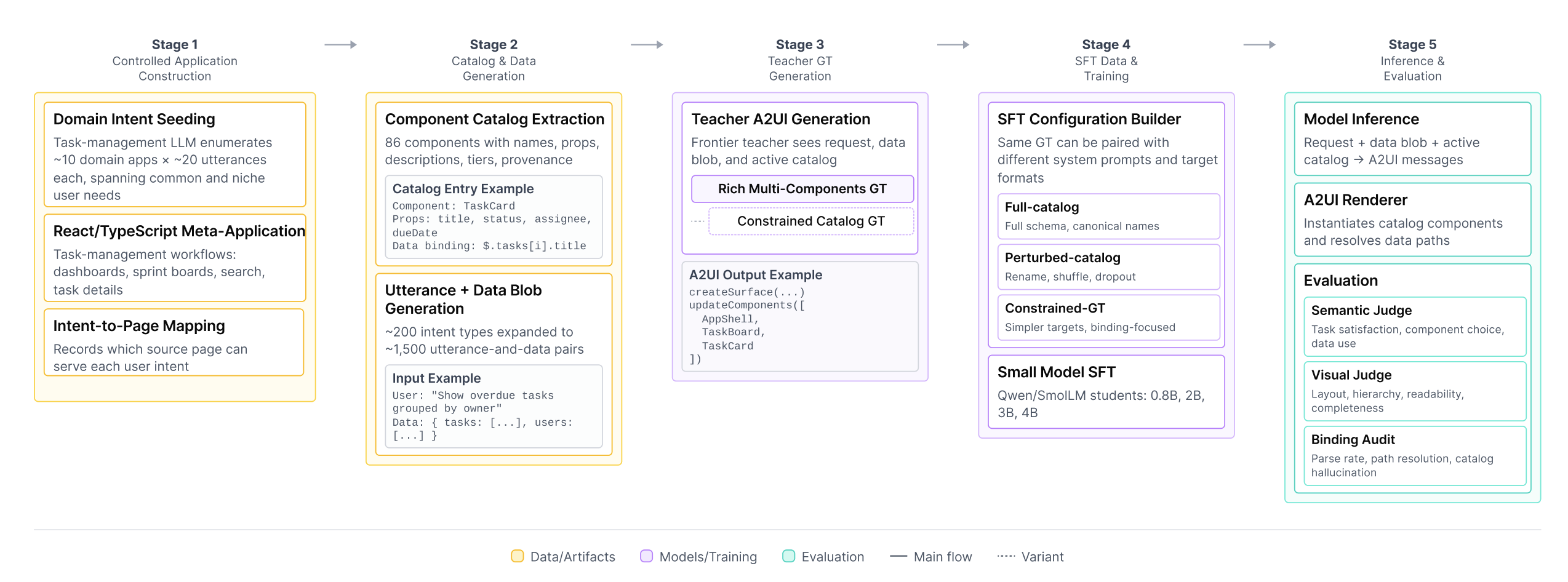}
\caption{Controlled pipeline for training small models to generate A2UI, from domain intent seeding and catalog extraction (Stages 1--2) through frontier-teacher ground truth (Stage 3), SFT under three configurations (Stage 4), and rendered inference with semantic, visual, and binding evaluation (Stage 5).}
\label{fig:pipeline}
\end{figure*}

\paragraph{Task and data.}
Given a user request, an A2UI component catalog, and a JSON data blob representing related data, the model generates A2UI messages that a client renderer instantiates into a live UI (Figure~\ref{fig:pipeline}). The protocol and catalog form the stable core of the pipeline, while the model-driven generation step is the adaptive part; the model's output must satisfy three contracts that give the system structure and guarantees over this inherently non-deterministic generation: syntactic validity, catalog validity, and data-binding validity (each \texttt{\{"path": ".../"\}} reference must resolve against the data blob). These three contracts define the contract-adherence metrics we report in \S\ref{sec:metrics}.

\paragraph{Data Schema example.}
The fragments in Figure~\ref{fig:pipeline} come from a single request; we trace it end-to-end here to make the data contract concrete. \textbf{(1) Utterance.} The user asks, e.g., ``Show overdue tasks grouped by owner.'' \textbf{(2) Data blob.} The request is paired with an application-data JSON object, e.g.\ \texttt{\{"tasks": [\{"title": "Fix login bug", "status": "overdue", "assignee": "u1", "dueDate": "2026-05-01"\}, ...], "users": [\{"id": "u1", "name": "Alice"\}, ...]\}}. \textbf{(3) Active catalog.} The system prompt lists the components available for this surface with their prop signatures, e.g.: \texttt{\{TaskCard [molecular]: title:BoundString!, status:BoundString!, assignee:BoundString, dueDate:BoundString\}}; a \texttt{!} marks a required prop and \texttt{BoundString} is a prop that may be filled either by a literal or by a data binding. \textbf{(4) Output A2UI messages.} The model emits two messages: \texttt{createSurface} (opens a surface and selects the catalog, e.g.\ \texttt{\{"createSurface": \{"surfaceId": "main", "catalogId": "id1"\}\}}) followed by \texttt{updateComponents}, which supplies the component tree. For a more detailed introduction to the A2UI format, we refer the reader to the A2UI protocol specification~\cite{a2ui2025}.

The component catalog is generated from an existing React code base for a given application (e.g., task management or cloud-ops management): our pipeline extracts reusable React components and converts them to an A2UI catalog. To generate training data, the pipeline combines real and synthesized user utterances with binding data for the application domain; a teacher model then produces the target A2UI message output given the utterance, data, and component catalog. The SFT data-construction strategies described in the next section are applied to this data to produce the training set.

We call these teacher outputs ground truth (GT) by convention, but they are \emph{one} valid solution, not a unique correct answer: many component selections and layouts can satisfy the same request. No metric compares a student's output to the teacher's---judges score the request-to-UI relationship, paths are checked against the data blob, and components against the active catalog---so valid divergence is never penalized; the cost falls on output diversity (\S\ref{sec:limitations}).

\begin{figure*}[t]
\centering
\begin{minipage}[t]{0.49\textwidth}
  \centering
  \includegraphics[width=\linewidth]{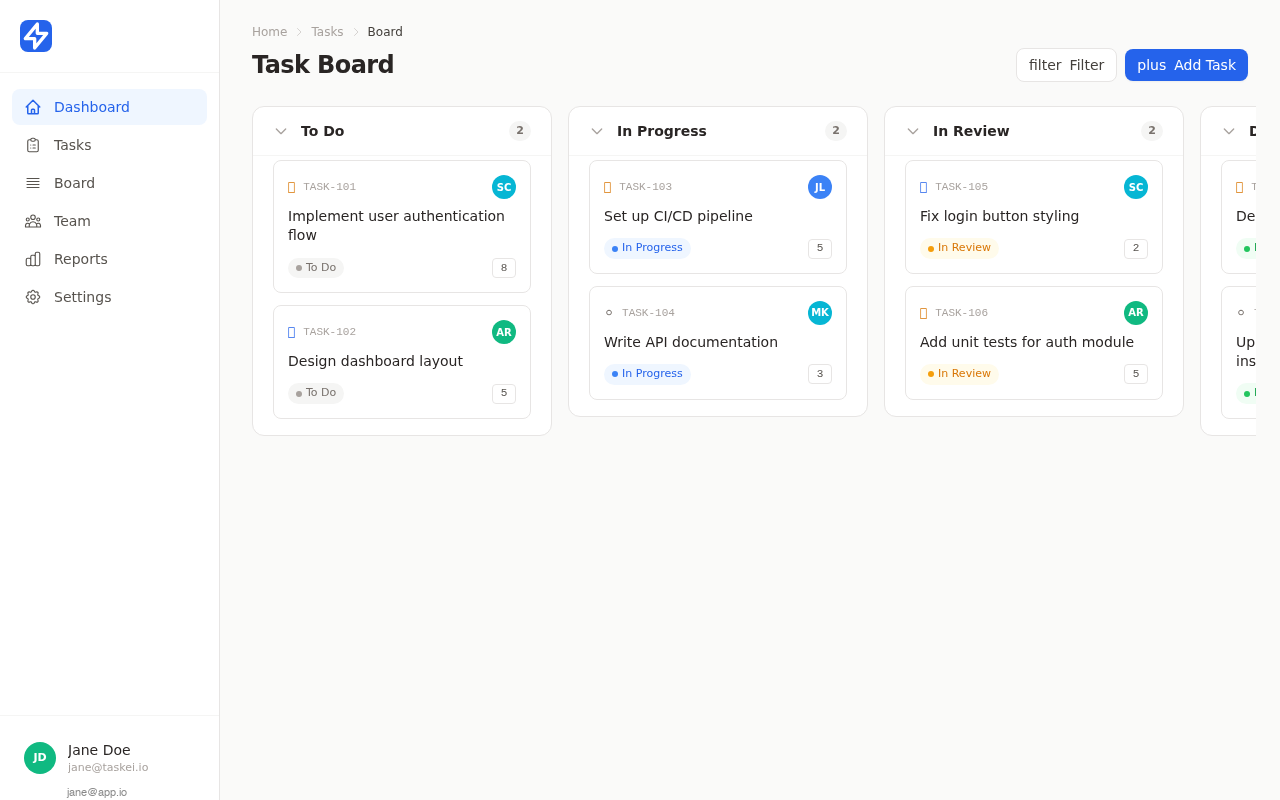}\\[3pt]
  {\footnotesize (a) ``Kanban view, group by status, cards with name, assignee, date, status indicator.''}
\end{minipage}\hfill
\begin{minipage}[t]{0.49\textwidth}
  \centering
  \includegraphics[width=\linewidth]{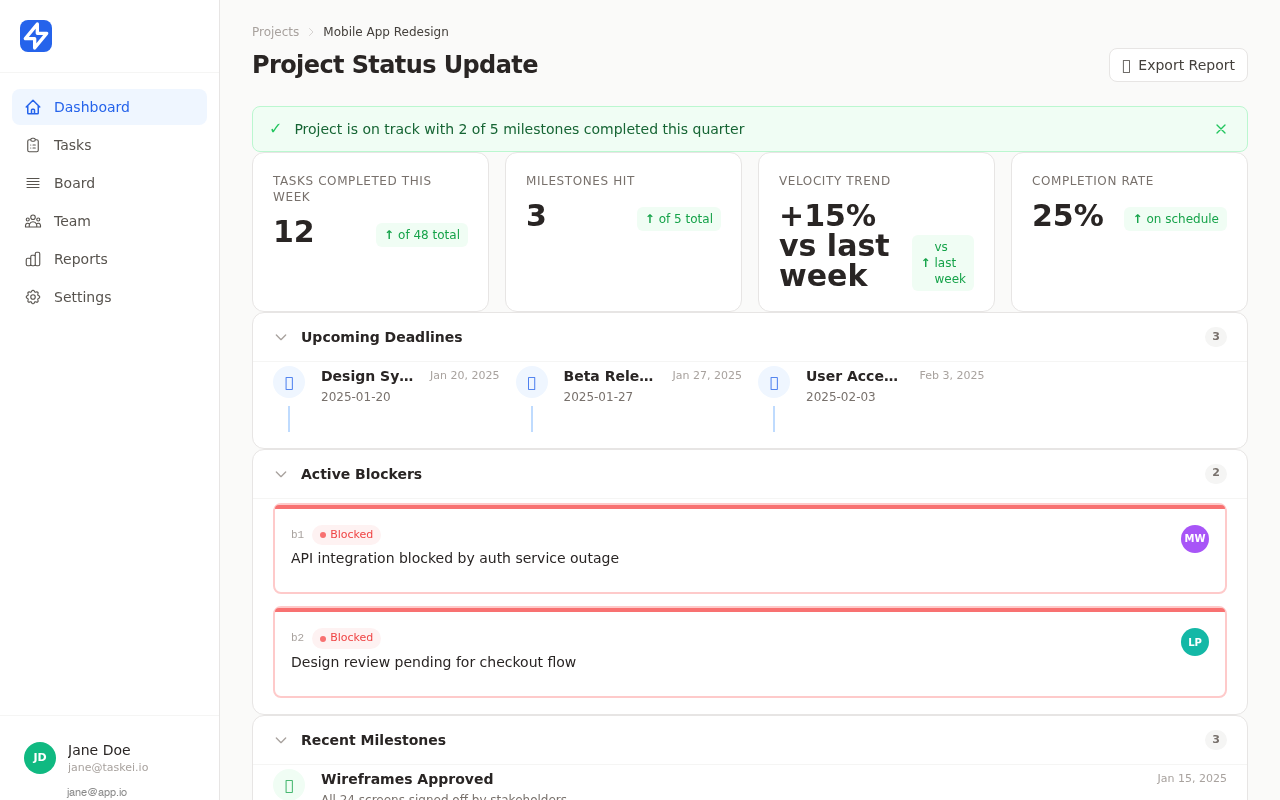}\\[3pt]
  {\footnotesize (b) ``Compose a project status update with auto-populated metrics: tasks completed, milestones, deadlines, blockers.''}
\end{minipage}
\caption{Rendered task-management outputs from the 4B Perturbed-catalog student on two representative in-distribution queries. Each surface is produced by a single forward pass through the 4B model and rendered client-side via extracted TSX component stubs with data-blob injection; the quoted text is the user request. These are the artifacts scored by the visual judge (\S\ref{sec:metrics}).}
\label{fig:screenshots}
\end{figure*}

We evaluate two domains: a task-management application (86-component catalog, $\sim$1{,}500 utterance-and-data pairs, $\sim$1{,}015 training / 320 ID (in distribution) test / 134 OOD (out of distribution) test) and a separate cloud-management console (compared in the cross-domain replication; \S\ref{sec:cross_domain}). For task management, the OOD test uses an expanded catalog condition containing 29 components never observed during training plus the core components needed to construct valid pages. This condition simulates catalog evolution---the component catalog changing after the model is trained---and tests which SFT strategy remains reliable as the system's stable component contract evolves, isolating the stable backbone (the A2UI protocol and core components the student must always honor) from the adaptive catalog surface that grows over time. Figure~\ref{fig:screenshots} shows two rendered task-management surfaces from our recommended 4B Perturbed-catalog student, to make concrete the kind of output the metrics below score.

\subsection{SFT Strategies}
\label{sec:sft_strategies}
We compare three SFT data-construction strategies. Full-catalog is the canonical baseline: every training and test sample sees the full 86-component catalog in the system prompt. Full-catalog ground truth A2UI messages are generated by instructing the teacher to build multi-level layouts with several component types rather than delegate to a single page-level component. Perturbed-catalog starts from the same rich multi-component ground truth but randomizes the catalog shown in each training sample's system prompt: component order is shuffled, unused components may be dropped, and 20\% of component names are renamed, with renames applied consistently to both the system prompt and the output A2UI message. This tests whether the student learns to read the active catalog rather than memorize a fixed component vocabulary. Constrained-GT keeps the full catalog in the system prompt and uses a similar but milder 5\% rename rate, but changes the teacher's target component number distribution: the teacher is prompted to produce simpler A2UI trees and a component-count limit. This strategy tests whether smaller, more regular targets makes it easier for small student models to learn from. All strategies share the same training queries.

\paragraph{SFT example format and prompt variants.}
Each SFT example has three logical fields: (1) a system prompt containing A2UI protocol rules and, depending on the configuration, a component catalog; (2) a user message containing the natural-language request and a \texttt{\textless data\textgreater} JSON blob; and (3) a target output containing A2UI messages. All catalog-conditioned prompts instruct the model to generate A2UI rather than React code, use only components present in the active catalog, construct a valid component tree, and bind data-dependent props through JSON path references into the provided data blob. The two-message target format is: \texttt{createSurface} (create the UI surface and select the catalog) and \texttt{updateComponents} (provide the component tree, component props, child relationships, and data bindings).

\paragraph{Ground-truth modes.}
Rich multi-component GT asks the teacher to build multi-level layouts using several component types rather than delegating to a single page-level component; this setting is referred to as ``deep-composition.'' It is the default for Full-catalog and Perturbed-catalog. Constrained-GT changes the target distribution: the teacher is encouraged to use simpler canonical page patterns and fewer components. On the task-management benchmark this can make outputs less semantically rich in-distribution; on the cloud-management benchmark the constrained targets remain medium-compact and do not show the same ceiling. The consistent benefit is stronger executable data binding under expanded-catalog OOD shift. The Perturbed-catalog rename augmentation is applied consistently to both the catalog shown in the system prompt and the component names in the target output. This is the correct construction for testing whether the model reads the catalog at inference time, but it also creates a recognition-vs-novelty effect: on the familiar training-domain catalog, the model may emit memorized renamed variants; on genuinely novel OOD catalogs, it tends to follow the presented names. Constrained-GT's milder 5\% rename rate is therefore less exposed to vocabulary pollution.

\subsection{Training and Inference Configuration}
\label{sec:training_config}
Students are LoRA-SFT'd \cite{hu2022lora} (rank $r{=}8$, learning rate $5{\times}10^{-5}$, 4 epochs, 8192-token context, per-device batch size 1 with gradient accumulation 8) from Qwen 3.5 base checkpoints \cite{qwen2026qwen35} at 0.8B, 2B, and 4B parameters, with next-token prediction on serialized A2UI assistant messages and Qwen thinking disabled for both training targets and inference. Training uses ${\sim}1{,}015$ filtered examples from the unified split; evaluation uses the 320 evaluable unified-test examples plus the expanded-catalog OOD sets. Unless otherwise stated, all SFT runs use the same train/test split, the same user requests and data blobs, and differ only in the prompt/target construction of \S\ref{sec:sft_strategies}. At evaluation time, students are decoded with the same no-think setting used during training. For the main two-message format, the model does not generate the application data blob; instead, the renderer receives the aligned input data blob and evaluates generated \texttt{\{"path": "..."\}} references against that blob. Main-body semantic and visual scores for Perturbed-catalog and Constrained-GT use deterministic component-name canonicalization.

\subsection{Metrics and Judges}
\label{sec:metrics}
We evaluate A2UI generation along four axes. The first three are contract-adherence metrics---one per contract from \S\ref{sec:methodology}---and the fourth measures rendered-output quality:
\textbf{(1) Parsing and rendering correctness} (parse rate, render success \%): does the model emit syntactically valid A2UI that the renderer can instantiate? Targets the syntactic-validity contract.
\textbf{(2) Catalog adherence} (bad-component / hallucination rate, in \%, computed by string-matching emitted component names against the active catalog): does the model use only components in the catalog? Targets the catalog-validity contract.
\textbf{(3) Path-binding correctness} (paths valid \%, computed by JSON-path traversal of the data blob against each \texttt{\{"path": ".../"\}} reference): does each binding actually resolve at render time? Targets the data-binding-validity contract; this is the only metric that measures whether the rendered UI will display the user's data rather than blanks.
\textbf{(4) Quality on the rendered output} (semantic and visual judge scores, 1--5). The semantic judge (Sonnet 4.6; \cite{anthropic2026sonnet46}) sees the user request and the A2UI message structure and rates whether the UI satisfies the request, selects appropriate components, and uses the data coherently. The visual judge (Opus 4.6; \cite{anthropic2026opus46}) sees the rendered screenshot and rates layout coherence, visual hierarchy, density, readability, and whether the page appears complete rather than broken or sparse.

\subsection{Judge Calibration}
\label{sec:judge_calibration}
The primary automatic scores use Claude-family judges: Sonnet 4.6~\cite{anthropic2026sonnet46} for semantic A2UI-message quality and Opus 4.6~\cite{anthropic2026opus46} for screenshot-based visual quality. Because some teacher and frontier-generator conditions also use Claude-family models, we add two calibration checks; they support relative comparisons and identify judge bias rather than justifying claims about absolute human preference.

\paragraph{Cross-family judge.}
We re-judge a stratified sample of up to 40 rows from each of 8 systems with an open-source multimodal judge, Qwen3-VL-235B-A22B \cite{qwen2025qwen3vl}, which scores both the A2UI message structure (semantic) and the rendered screenshot (visual). After filtering unmatched or missing rows, the paired comparison contains 308 judged items. The goal is to test whether \textbf{model rankings}, which are the basis of most comparative claims in the paper, are preserved under a different judge family.

\begin{table}[h]
\centering
\caption{Cross-family calibration: Qwen3-VL vs.\ the primary judges.}
\label{tab:cross_judge}
\small
\begin{tabular}{lcc}
\toprule
\textbf{Metric} & \textbf{Semantic} & \textbf{Visual} \\
\midrule
Per-model-mean Spearman $\rho$ & \textbf{0.98} & \textbf{0.96} \\
Per-item Spearman $\rho$ & 0.59 & 0.86 \\
Mean offset (Qwen3-VL $-$ base) & +1.31 & +0.19 \\
Within $\pm$1 agreement & 63\% & 97\% \\
\bottomrule
\end{tabular}
\end{table}

The per-model rank correlation (Table~\ref{tab:cross_judge}) is 0.98 for semantic and 0.96 for visual quality, supporting broad relative claims: larger students beat smaller ones on the curated strategies, frontier/teacher systems remain strong reference points, and Perturbed-catalog scores above Full-catalog under OOD catalog shift. We do not treat it as evidence for small cell-level differences, which still require paired tests. Qwen3-VL assigns higher semantic scores on average (+1.31), so we use it only as a rank-consistency check; on the visual axis it tracks the primary judge closely (+0.19, 97\% within $\pm$1) and independently rates Perturbed-catalog's in-distribution visual lowest, supporting the diagnosis that renamed component names cause renderer-visible failures rather than a single-judge artifact.

\paragraph{Human annotation study.}
Three team members independently rated 50 stratified 4B Full-catalog outputs (150 ratings) on three 1--5 dimensions: visual quality, task satisfaction, and component appropriateness. Samples were presented in random order with the generating configuration hidden; annotators saw the user request, the rendered screenshot, and the model-generated code. The three dimensions decompose the two automatic axes rather than adding new ones: visual quality is matched against the visual judge, while task satisfaction and component appropriateness both map onto the semantic judge, which bundles those two sub-judgments.

\begin{table}[h]
\centering
\caption{Human alignment (Spearman $\rho$, $n{=}50$ items, three annotators). Judge--Human is the mean correlation between the LLM judge and each annotator individually; Human--Human is the mean pairwise inter-annotator correlation. Both are single-rater correlations and therefore directly comparable.}
\label{tab:human_study}
\small
\setlength{\tabcolsep}{5pt}
\begin{tabular}{lcc}
\toprule
\textbf{Dimension} & \textbf{Judge--Human} & \textbf{Human--Human} \\
\midrule
Visual      & 0.48 & 0.46 \\
Task sat.\  & 0.25 & 0.35 \\
Component   & 0.08 & 0.18 \\
\bottomrule
\end{tabular}
\end{table}

Visual quality is the most reliably judged axis (Table~\ref{tab:human_study}): the judge agrees with an individual annotator about as well as annotators agree with each other ($0.48$ vs.\ $0.46$). The two semantic sub-dimensions have weaker positive correlation. The judge tracks task satisfaction less closely than annotators track each other ($0.25$ vs.\ $0.35$). While for component appropriateness (judging if the component subset selection is optimal, which is a hard combinatorial optimization problem) even human agreement is itself low ($0.18$). 

%% file: sections/evaluation.tex
\section{Results}

We organize the experiments around the three dials introduced in \S1---SFT strategy, model size, and component-catalog size---followed by the cross-domain replication of RQ4. All results use the renderer-based pipeline of \S\ref{sec:methodology}.

\subsection{RQ1: Which SFT Strategy Wins, and on Which Axis?}
\label{sec:rq1}

\begin{table}[!tbp]
\centering
\caption{4B strategy comparison on ID and OOD tests (1--5 LLM-judge scores; rest in \%). Semantic/Visual values show $\pm$ 95\% bootstrap CI half-widths over test items (\S\ref{sec:significance}); pairwise differences are tested with paired Wilcoxon signed-rank.}
\label{tab:strategy_id_ood}
\scriptsize
\setlength{\tabcolsep}{3pt}
\begin{tabular}{llccccc}
\toprule
\textbf{Setting} & \textbf{Strategy} & \textbf{Semantic} & \textbf{Visual} & \textbf{Parsed} & \textbf{Paths valid} & \textbf{Bad comp.} \\
\midrule
\multirow{3}{*}{In-dist.}
& Full-catalog & 3.01\,\tiny{$\pm$.09} & 2.78\,\tiny{$\pm$.10} & 98.4\% & 50.8\% & \textbf{1.9\%} \\
& Perturbed-catalog & \textbf{3.45}\,\tiny{$\pm$.09} & \textbf{3.09}\,\tiny{$\pm$.10} & 99.1\% & 92.6\% & 7.8\% \\
& Constrained-GT & 2.37\,\tiny{$\pm$.07} & 2.97\,\tiny{$\pm$.09} & 98.7\% & \textbf{93.3\%} & 0.2\% \\
\midrule
\multirow{3}{*}{OOD}
& Full-catalog & 2.67\,\tiny{$\pm$.12} & 2.98\,\tiny{$\pm$.16} & 95.5\% & 56.4\% & 9.7\% \\
& Perturbed-catalog & \textbf{3.16}\,\tiny{$\pm$.14} & 3.21\,\tiny{$\pm$.18} & \textbf{99.3\%} & 87.7\% & \textbf{0.7\%} \\
& Constrained-GT & 2.76\,\tiny{$\pm$.14} & \textbf{3.32}\,\tiny{$\pm$.20} & 98.5\% & \textbf{98.1\%} & 0.5\% \\
\bottomrule
\end{tabular}
\end{table}

\paragraph{Augmentation vs.\ baseline.} As shown in Table~\ref{tab:strategy_id_ood}, the Full-catalog baseline lags behind the two augmented strategies on almost all metrics, especially on data-binding correctness, which requires generalization rather than pure memorization: Full-catalog resolves only $50.8\%$ of paths ID and $56.4\%$ OOD, whereas the augmented strategies reach $92.6$--$93.3\%$ ID and $87.7$--$98.1\%$ OOD. The gap also shows on judged quality (e.g.\ ID semantic $3.01$ for Full-catalog vs.\ $3.45$ for Perturbed-catalog) and widens under catalog shift, consistent with the baseline leaning on memorization rather than generalization.

\paragraph{Two augmented strategies specialize on different axes.} Perturbed-catalog wins judged semantic quality both ID and OOD and posts the lowest OOD bad-component rate (0.7\%). Constrained-GT wins executable binding (OOD path resolution 98.1\% vs.\ 87.7\%, a $10.4$\% absolute gain that drops the residual unresolved-path rate from $12.3\%$ to $1.9\%$, a $\sim$$6.5\times$ reduction) and OOD visual score (3.32 vs.\ 3.21).

\paragraph{Statistical significance of strategy differences.}
\label{sec:significance}
We test the 4B strategy comparisons in Table~\ref{tab:strategy_id_ood} for statistical significance. Because all strategies are scored on the same test items, we use paired tests on per-item score differences: the Wilcoxon signed-rank test (normal approximation with tie and continuity correction) and a paired bootstrap 95\% CI on the mean per-item difference (2{,}000 resamples). These quantify test-item sampling uncertainty.

To check that strategy differences are not artifacts of a single training run, we additionally retrained the Full-catalog 4B student from scratch across five random seeds (with greedy decoding at inference, so the training seed is the only source of variance). The training-seed standard deviation is small on every axis---$\pm0.05$ semantic and $\pm0.03$ visual in-distribution, $\pm0.05$ semantic and $\pm0.02$ visual OOD---far below the strategy gaps in Table~\ref{tab:strategy_id_ood} (e.g.\ Perturbed-catalog${-}$Full-catalog ${=}{+}0.45$ semantic ID). The strategy differences we report are therefore not attributable to training-seed noise.

\begin{table}[h]
\centering
\caption{Paired Wilcoxon signed-rank tests on per-item differences (4B), semantic and visual. $\Delta$ is the mean per-item difference (row strategy minus column strategy). \textbf{Pert.}, \textbf{Cons.}, and \textbf{Full} abbreviate Perturbed-catalog, Constrained-GT, and Full-catalog.}
\label{tab:significance}
\small
\setlength{\tabcolsep}{4pt}
\begin{tabular}{llcccc}
\toprule
& & \multicolumn{2}{c}{\textbf{Semantic}} & \multicolumn{2}{c}{\textbf{Visual}} \\
\cmidrule(lr){3-4} \cmidrule(lr){5-6}
\textbf{Setting} & \textbf{Pair} & \textbf{$\Delta$} & \textbf{$p$} & \textbf{$\Delta$} & \textbf{$p$} \\
\midrule
\multirow{3}{*}{ID}
& Pert.\ vs Full   & $+0.45$ & ${<}10^{-10}$ & $+0.34$ & ${<}10^{-6}$ \\
& Pert.\ vs Cons.  & $+1.09$ & ${<}10^{-12}$ & $+0.14$ & $0.04$ \\
& Cons.\ vs Full   & $-0.65$ & ${<}10^{-12}$ & $+0.19$ & $0.009$ \\
\midrule
\multirow{3}{*}{OOD}
& Pert.\ vs Full   & $+0.52$ & ${<}10^{-5}$ & $+0.23$ & $0.02$ \\
& Pert.\ vs Cons.  & $+0.43$ & ${<}10^{-4}$ & $-0.12$ & $0.23$ \\
& Cons.\ vs Full   & $+0.09$ & $0.36$ & $+0.35$ & $0.001$ \\
\bottomrule
\end{tabular}
\end{table}

Table~\ref{tab:significance} confirms Perturbed-catalog's semantic advantage over Full-catalog is significant both ID and OOD, and Constrained-GT's OOD visual advantage over Full-catalog is significant. The only non-significant comparisons are Perturbed-catalog vs.\ Constrained-GT on OOD visual ($p{=}0.23$, both strong on that axis) and Constrained-GT vs.\ Full-catalog on OOD semantic ($p{=}0.36$); the latter reflects that these strategies occupy different operating points (binding vs.\ baseline) rather than a strict quality ordering, consistent with our ``no single strategy dominates'' framing.

\subsection{RQ2: Model Size, Cost, and Quality Trade-Offs}
\label{sec:rq2}

\begin{figure}[!tbp]
\centering
\includegraphics[width=\linewidth]{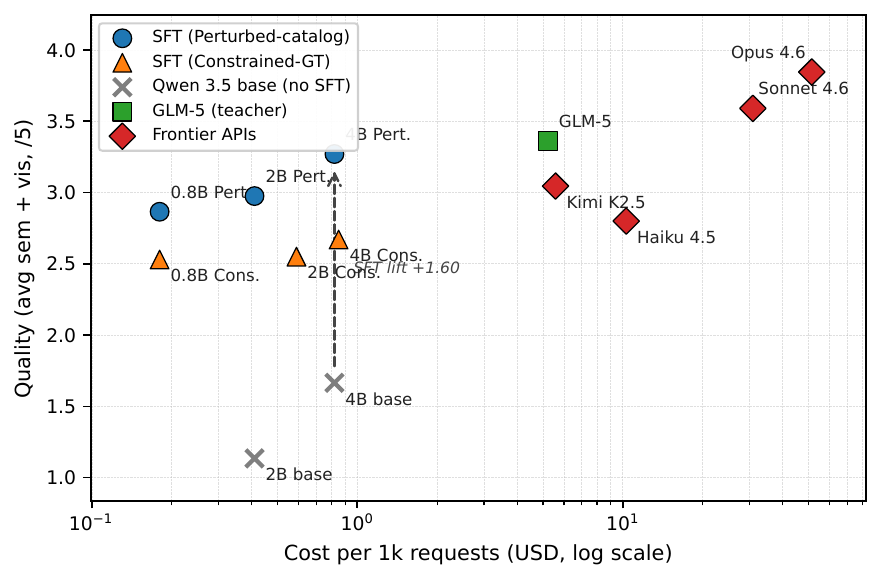}
\caption{Cost vs.\ quality across SFT students (Qwen 3.5 0.8B/2B/4B) and frontier APIs at the in-distribution average workload. Quality is the average of semantic and visual judge scores (1--5); cost is \$/1k queries. The 4B Perturbed-catalog student sits on the Pareto front. Workload and pricing derivation are detailed in the text.}
\label{fig:cost_quality}
\end{figure}

\begin{table}[!tbp]
\centering
\caption{Semantic and visual quality across model sizes on ID and OOD tests. Strategy ranking is mostly preserved at every size. \textbf{Full}, \textbf{Pert.}, and \textbf{Cons.} abbreviate Full-catalog, Perturbed-catalog, and Constrained-GT.}
\label{tab:model_scale}
\scriptsize
\setlength{\tabcolsep}{4pt}
\begin{tabular}{lcccccc}
\toprule
& \multicolumn{3}{c}{\textbf{In-distribution}} & \multicolumn{3}{c}{\textbf{OOD}} \\
\cmidrule(lr){2-4} \cmidrule(lr){5-7}
\textbf{Size} & \textbf{Full} & \textbf{Pert.} & \textbf{Cons.} & \textbf{Full} & \textbf{Pert.} & \textbf{Cons.} \\
\midrule
\multicolumn{7}{l}{\emph{Semantic}} \\
4B   & 3.01 & \textbf{3.45} & 2.37 & 2.67 & \textbf{3.16} & 2.76 \\
2B   & 2.78 & 3.02 & 2.32 & 2.27 & 2.87 & 2.63 \\
0.8B & 2.61 & 2.89 & 2.23 & 1.99 & 2.77 & 2.46 \\
\midrule
\multicolumn{7}{l}{\emph{Visual}} \\
4B   & 2.78 & \textbf{3.09} & 2.97 & 2.98 & 3.21 & \textbf{3.32} \\
2B   & 2.73 & \textbf{2.93} & 2.78 & 2.23 & 2.69 & \textbf{3.13} \\
0.8B & 2.64 & \textbf{2.84} & 2.83 & 2.47 & \textbf{2.63} & 2.33 \\
\bottomrule
\end{tabular}
\end{table}

\paragraph{Smoother degradation on ID tests; sharper degradation on OOD tests.}
Quality decreases monotonically with size on all strategies, with the 4B-to-0.8B gap in general larger on OOD test than on ID. On Full-catalog, semantic drops $0.40$ ID ($3.01 \to 2.61$) but $0.68$ OOD ($2.67 \to 1.99$); visual drops $0.14$ ID ($2.78 \to 2.64$) but $0.51$ OOD ($2.98 \to 2.47$). Constrained-GT shows the same trend, sharpest on visual ($0.14$ ID gap vs.\ $0.99$ OOD gap). The pattern can be explained by memorization-vs-generalization: small models lean on memorization of the training distribution and hold up reasonably well in distribution, but performance starts to degrade as the catalog starts to drift from training data. Practical implication: for ID (in distribution)-heavy or stable-catalog deployments, 0.8B could be considered shippable; for OOD-heavy or visually critical workloads, the 4B tier is a better fit. Constrained-GT sits 0.6--1.1 points below Perturbed-catalog on task-management ID semantic, as its component budget collides with this benchmark's page-level queries; \S\ref{sec:cross_domain} reports a domain where it instead matches Perturbed-catalog at 3.42 ID semantic.

\paragraph{4B SFT vs.\ frontier.} Figure~\ref{fig:cost_quality} places 0.8B/2B/4B Perturbed-catalog and Constrained-GT students against Claude Haiku 4.5, Sonnet 4.6, Opus 4.6 \cite{anthropic2025haiku45,anthropic2026sonnet46,anthropic2026opus46}, Kimi K2.5 \cite{moonshot2026kimi25}, and the GLM-5 teacher \cite{zai2026glm5} on cost vs.\ quality. The comparison is at the in-distribution average workload ($\sim$1{,}200 output tokens, mean across SFT recipes; the 2B Constrained-GT model is an outlier at $\sim$1{,}715 output tokens, with proportionally higher latency and cost). SFT \$/1k is amortized H100 cost at vLLM-class continuous-batching throughput; frontier \$/1k uses public API list pricing at the workload (Opus 4.6 \$5/M$+$\$25/M $\to$ \$51.50/1k; Sonnet 4.6 \$3/M$+$\$15/M $\to$ \$30.90/1k; Haiku 4.5 \$1/M$+$\$5/M $\to$ \$10.30/1k; Kimi K2.5 \$0.60/M$+$\$2.50/M $\to$ \$5.58/1k; GLM-5 via Z.ai $\to$ \$5.22/1k). The 4B Perturbed-catalog student sits on the Pareto front: it recovers $\sim$98\% of teacher semantic and $\sim$97\% of teacher visual quality, runs at 2--3$\times$ lower latency than Sonnet/Opus, and costs more than an order of magnitude less per-query than calling Claude through an API. 4B SFT also dominates Kimi K2.5 on both axes: higher quality (3.27 vs.\ 3.05 average) at 6.8$\times$ lower cost (\$0.82 vs.\ \$5.58/1k). The 0.8B tier deepens the gap on cost: $\sim$3.4\,s end-to-end latency on a single H100 at $\sim$\$0.18/1k. Sonnet and Opus remain ahead on absolute quality but at higher cost and slower latency.

\begin{figure}[h]
\centering
\includegraphics[width=\linewidth]{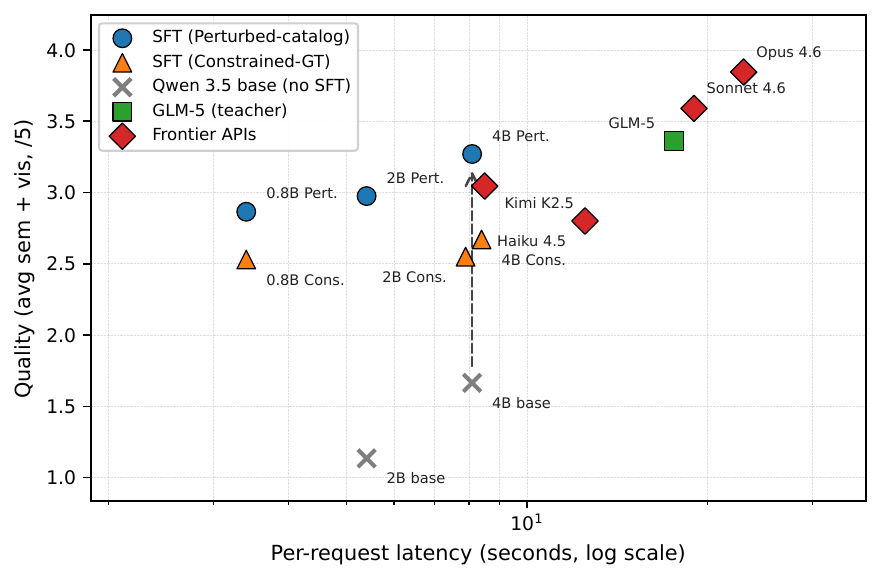}
\caption{Latency vs.\ quality variant of Figure~\ref{fig:cost_quality}, log-scale latency axis.}
\label{fig:latency_quality}
\end{figure}

\paragraph{Cost vs.\ quality on a latency axis.}
\label{sec:cost_latency}
Figure~\ref{fig:latency_quality} reports the same model set as Figure~\ref{fig:cost_quality} on a latency axis (per-request seconds, log scale) instead of cost. The same Pareto-front structure largely obtains: 4B Perturbed-catalog at $\sim$8.1\,s is faster than Haiku 4.5 ($\sim$12.5\,s), Kimi K2.5 ($\sim$8.5\,s), Sonnet 4.6 ($\sim$19\,s), and Opus 4.6 ($\sim$23\,s), and reaches higher quality than Haiku 4.5 and Kimi K2.5; the 0.8B SFT student is the fastest deployable point at $\sim$3.4 s. Sonnet/Opus reach higher absolute quality but at 19--23\,s per request, exceeding the latency budget most production dynamic-UI surfaces require. Pre-SFT base Qwen 3.5 (gray crosses, no SFT) is included to make the SFT lift visible at fixed inference cost: the base 4B model produces unusable A2UI (averaged sem $+$ vis $=$ 1.66; only 62\% parse rate) and SFT lifts the same 4B checkpoint to 3.27 average quality (Figure~\ref{fig:cost_quality} dashed arrow).

\subsection{RQ3: How Does Catalog Size Shape Student Quality?}
\label{sec:rq3}

\begin{table}[!tbp]
\centering
\caption{Catalog-size sweep across student sizes.}
\label{tab:catalog_size}
\small
\setlength{\tabcolsep}{4pt}
\begin{tabular}{lccc@{\hskip 10pt}ccc}
\toprule
& \multicolumn{3}{c}{\textbf{Semantic}} & \multicolumn{3}{c}{\textbf{Visual}} \\
\cmidrule(lr){2-4} \cmidrule(lr){5-7}
\textbf{Catalog} & \textbf{0.8B} & \textbf{2B} & \textbf{4B} & \textbf{0.8B} & \textbf{2B} & \textbf{4B} \\
\midrule
10  & 2.29 & 2.50 & 2.68 & 1.71 & 2.30 & 2.36 \\
15  & 2.35 & 2.56 & 2.67 & 1.96 & 2.55 & 2.59 \\
25  & 2.44 & 2.77 & 2.95 & 2.05 & 2.69 & 2.58 \\
50  & 2.39 & 2.76 & 2.88 & 2.03 & 2.75 & 2.80 \\
86  & \textbf{2.97} & \textbf{3.16} & \textbf{3.47} & \textbf{2.97} & \textbf{3.00} & \textbf{3.07} \\
\bottomrule
\end{tabular}
\end{table}

\paragraph{Quality scales monotonically with catalog size.} From 10 to 86 components (Table~\ref{tab:catalog_size}), both semantic and visual quality improve monotonically at all three student sizes (0.8B semantic $2.29\to2.97$; 2B $2.50\to3.16$; 4B $2.68\to3.47$). Semantic shows a critical-mass jump at C86 (4B: $2.95$ at C25 $\to 3.47$ at C86); visual scales smoothly. The model-size ordering 4B $>$ 2B $>$ 0.8B holds at every catalog size. Practical implication: do not artificially cap the catalog ``for small models''---a richer catalog is part of the student's effective output vocabulary. After SFT, even the 0.8B model benefits most from the full catalog (its semantic and visual both reach $2.97$ at C86), capable of utilizing a component catalog of close to ${\sim}100$ components.

\subsection{RQ4: Can We Replicate the Findings on Other Student Models/Domains?}
\label{sec:cross_domain}

To test whether the strategy ranking from the main task-management benchmark is a property of one base model and one application domain, we replicated the same three-strategy SFT protocol on a different base model (SmolLM 3B) and a different UI domain (a cloud-management console with a 47-component catalog including pages, controls, charts, and resource cards). The new domain uses an independent component catalog and an independent test split (74 ID examples / 50 OOD examples spanning held-out billing- and IAM-related components). For OOD, the system prompt explicitly mandates the held-out components per query (oracle directive); this differs from the main-benchmark OOD test, which is unaided. Note that this replication varies base model and domain together, so it tests whether the strategy ranking survives a joint change of both; it cannot attribute the resulting differences to either factor alone.

\begin{table}[h]
\centering
\caption{Cross-domain replication on SmolLM 3B trained on the cloud-management UI domain.}
\label{tab:smol_replication}
\scriptsize
\begin{tabular}{lcccccc}
\toprule
\textbf{Setting} & \textbf{Strategy} & \textbf{Bind\%} & \textbf{Halluc\%} & \textbf{Sem} & \textbf{Vis} \\
\midrule
\multirow{3}{*}{In-dist.}
& Full-catalog       & 91.8 & 0.06 & 3.54 & 3.33 \\
& Perturbed-catalog  & 92.4 & \textbf{0.00} & \textbf{3.70} & 3.41 \\
& Constrained-GT     & \textbf{96.6} & \textbf{0.00} & 3.42 & \textbf{3.77} \\
\midrule
\multirow{3}{*}{OOD}
& Full-catalog       & 88.7 & \textbf{0.00} & 3.11 & 2.87 \\
& Perturbed-catalog  & 85.6 & 0.76 & 3.08 & 2.96 \\
& Constrained-GT     & \textbf{98.6} & \textbf{0.00} & \textbf{3.17} & \textbf{3.22} \\
\bottomrule
\end{tabular}
\end{table}

The SFT recipe findings replicate on this different base model and UI domain (Table~\ref{tab:smol_replication}): the augmented strategies again define stronger quality and binding operating points than the unaugmented baseline. Perturbed-catalog again leads ID semantic (3.70 vs.\ 3.54 for Full-catalog), and Constrained-GT again wins executable binding (98.6\% OOD path resolution vs.\ 85.6--88.7\% for the others). This mirrors its 98.1\% on task-management, confirming the binding specialization is not domain-specific. Constrained-GT additionally wins visual coherence both ID (3.77) and OOD (3.22) on this cleaner-catalog domain. Both augmented strategies beat Full-catalog on the main cloud-management axes (ID semantic, ID visual, OOD semantic, OOD visual, and OOD path resolution), with the only exception being a small tie on ID hallucination (where all three strategies round to $\leq$0.06\%).

\paragraph{Constrained-GT semantic on this domain matches Perturbed-catalog.} Constrained-GT ID semantic on the cloud-management domain is 3.42, OOD 3.17, well above the task-management Constrained-GT numbers (2.23--2.37 ID, 2.46--2.76 OOD; Table~\ref{tab:model_scale}). We attribute the task-management gap to the strategy's component budget colliding with that benchmark's page-level test queries (sprint boards, dashboards, project summaries that need rich skeleton components); the cloud-management domain has fewer queries that exceed the budget. The lesson for practitioners is that Constrained-GT semantic quality depends on whether the test workload's natural component count fits inside the strategy's budget, not on a property of the strategy itself. On task-management, Constrained-GT posts 98.1\% OOD path resolution; on the cloud-management domain, 98.6\% (vs.\ 85--89\% for the other strategies in either domain), so its specialization for catalog-adherence and binding-executability is robust across base models.

%% file: sections/discussion.tex
\section{Discussion and Limitations}
\label{sec:limitations}

\paragraph{What transfers beyond A2UI.}
The three contracts we evaluate---syntactic validity, catalog membership, resolvable bindings---exist in any protocol where a model selects catalog components and binds props to data, and our mechanisms are correspondingly generic: perturbation forces the student to read the active catalog rather than memorize a vocabulary, the quality-versus-binding split follows from what the targets emphasize, and catalog richness helps because the catalog is the student's effective output vocabulary. The two-message envelope, the path syntax, and the catalog's prompt-token cost are A2UI-specific, and shift absolute cost and latency but not rankings. Grammar-constrained decoding would remove most of SFT's parse-rate advantage, and renderer-side binding resolution would neutralize the axis on which Constrained-GT wins; we have not tested other protocols.

\paragraph{Limitations.}
Each record pairs one request with one A2UI response, whereas deployments often involve multi-turn updates. Each target is also a single teacher generation, so students inherit that teacher's layout idioms; sampling several solutions per request is a natural next step. Two domains cannot establish generality: both are real React/TypeScript applications chosen to differ on the axis that mattered---catalog size (86 vs.\ 47) and page character (board/list-heavy vs.\ metric/table-heavy)---but consumer, data-visualization, and forms-heavy surfaces may differ. Our largest catalog (86) exceeds typical A2UI catalogs, though a general-purpose model spanning many domains would need substantially larger ones, and more succinct message encodings than A2UI could further reduce latency. Finally, our study calibrates the judges on 50 items rated by three team members and is not a usability evaluation; a larger, fully blind study with external annotators would give a more reliable estimate, as the per-annotator spread in \S\ref{sec:judge_calibration} indicates, and showing surfaces are usable would require task-based user testing.

%% file: sections/conclusion.tex
\section{Conclusion}

We systematically studied three controllable design choices for training small models to generate declarative UI via the A2UI protocol: SFT data-construction strategy, model size, and component-catalog size. Overall, across two base models (Qwen 3.5 4B; SmolLM 3B) and two UI domains, augmented data construction consistently creates more robust and higher quality UI than the canonical Full-catalog baseline.

Taking the three dimensions together: catalog size scales student quality monotonically, so a richer component catalog is preferable whenever available. For model size and data-construction strategy, we recommend Perturbed-catalog at 4B as the default for general dynamic-UI generation; 0.8B is viable when low latency is a hard requirement and the queries are mostly in-distribution with a simpler catalog. Constrained-GT is preferable when data-binding rate on a dynamic catalog is the top consideration.
These recommendations rest on single-turn evaluation over two domains and on automatic judges rather than usability testing (\S\ref{sec:limitations}).